\documentclass[a4paper,11pt]{article}
\usepackage{pos}
\usepackage{subcaption}

\newcommand{\de}{\ensuremath{\delta} }

\newcommand{\La}{\ensuremath{\Lambda} }

\newcommand{\cN}{\ensuremath{\mathcal N} }

\newcommand{\ket}[1]{\ensuremath{\left| #1 \right\rangle} }

\newcommand{\matElem}[2]{\ensuremath{\left| #1 \right\rangle \left\langle #2 \right|}} 

\newcommand{\eq}[1]{Eq.~\ref{#1}}
\newcommand{\refcite}[1]{Ref.~\cite{#1}}

\newcommand{\ahat}{\ensuremath{\hat a}}
\newcommand{\ahatdag}{\ensuremath{\hat{a}^{\dagger}}}

\usepackage{mleftright,xparse}
\NewDocumentCommand\xDeclarePairedDelimiter{mmm}
{%
	\NewDocumentCommand#1{som}{%
		\IfNoValueTF{##2}
		{\IfBooleanTF{##1}{#2##3#3}{\mleft#2##3\mright#3}}
		{\mathopen{\csname##2\endcsname#2}##3\mathclose{\csname##2\endcsname#3}}%
	}%
}
\xDeclarePairedDelimiter{\av}{\langle}{\rangle}
\xDeclarePairedDelimiter{\abs}{\lvert}{\rvert}
\NewDocumentCommand\braket{somm}{%
	\IfNoValueTF{#2}{\mleft\langle #3\,|#4\mright\rangle}{NOTIMPLEMENTED}
}
\NewDocumentCommand\opbraket{sommm}{%
	\IfNoValueTF{#2}
	{\IfBooleanTF{#1}{\langle#3|#4|#5\rangle}{\mleft\langle #3 \left| #4 \right| #5 \mright\rangle}}
	{\mathopen{\csname#2\endcsname\langle}#3\mathopen{\csname#2\endcsname|} #4 \mathclose{\csname#2\endcsname|} #5\mathclose{\csname#2\endcsname\rangle}}
}

\title{Quantum Computing for the Wess--Zumino Model}

\author*{Christopher Culver}
\author{David Schaich}

\affiliation{Department of Mathematical Sciences, University of Liverpool, Liverpool L69 7ZL, United Kingdom}

\emailAdd{C.Culver@liverpool.ac.uk}
\emailAdd{David.Schaich@liverpool.ac.uk}

\abstract{Future quantum computers will enable novel sign-problem-free studies of dynamical phenomena in non-perturbative quantum field theories, including real-time evolution and spontaneous supersymmetry breaking.  We are investigating applications of quantum computing to low-dimensional supersymmetric lattice systems that can serve as testbeds for existing and near-future quantum devices.  Here we present initial results for the $\cN = 1$ Wess--Zumino model in 1+1~dimensions, building on our prior analyses of 0+1-dimensional supersymmetric quantum mechanics.  In addition to exploring supersymmetry breaking using the variational quantum eigensolver, we consider the prospects for real-time evolution.}

\FullConference{%
  The 38th International Symposium on Lattice Field Theory, LATTICE2022\\
  8--13 August 2022\\
  Bonn, Germany
}

\begin{document}
\maketitle

\section{Introduction}\label{sec:intro}
Supersymmetry is an extension of Poincar\'e symmetry that has many important applications throughout theoretical physics.
These include potential extensions of the standard model, insight into fundamental properties of quantum field theory (QFT), and holographic dualities with theories of quantum gravity.
Spontaneous symmetry breaking is an important topic in each of these realms.
In particular, dynamical \emph{super}symmetry breaking is a requirement of any experimentally viable supersymmetric model of new physics, since experiments have not yet discovered superpartners of the known particles of the standard model.
Supersymmetry breaking is also a feature of simpler QFTs, which we consider in this proceedings.

To non-perturbatively analyze supersymmetric QFTs, we employ lattice regularization.
While Monte Carlo importance sampling studies of supersymmetric lattice QFTs have been performed for many years (see Refs.~\cite{Kadoh:2016eju, Bergner:2016sbv, Schaich:2022xgy} for recent reviews), sign problems can prevent this approach from considering key dynamical phenomena including real-time evolution and spontaneous supersymmetry breaking~\cite{Bergner:2016sbv, Schaich:2022xgy}.
Quantum computing in principle provides a novel means to study these phenomena without introducing sign problems.

Existing and near-future quantum devices feature modest numbers (tens to hundreds) of qubits with relatively high error rates, widely described as Noisy Intermediate-Scale Quantum (NISQ) technology~\cite{preskill2018quantum}.
Lattice field theory studies employing such NISQ hardware are limited to small systems and shallow circuit depths, leaving the calculations within the reach of classical diagonalization.
Even in the absence of quantum advantage, studies of these small systems are important to explore, test, verify and refine quantum algorithms as hardware capabilities continue to improve~\cite{Alexeev:2020xrq}.

Here we investigate the $\cN = 1$ Wess--Zumino model in 1+1~dimensions, building on our prior analyses of 0+1-dimensional supersymmetric quantum mechanics~\cite{Culver:2021rxo}.
This is arguably the simplest supersymmetric quantum field theory, and has previously been the subject of lattice investigations from a variety of approaches.
In addition to lattice calculations employing the traditional Lagrangian formulation~\cite{Catterall:2003ae, Wozar:2011gu}, other studies also consider the continuous-time Hamiltonian formulation~\cite{Beccaria:2001qm, Beccaria:2003gt, Beccaria:2004pa, Beccaria:2004ds}, the fermion loop formulation~\cite{Steinhauer:2014yaa}, and tensor network formulations~\cite{Kadoh:2018hqq, Meurice:2020pxc}.
See \refcite{Kadoh:2016eju} for a brief review.

Our current focus is on dynamical supersymmetry breaking in the 1+1d Wess--Zumino model for specific prepotentials to be discussed below.
We will use the variational quantum eigensolver~(VQE) to explore this.
In the next section we begin by briefly summarizing the model, then in Section~\ref{sec:qc} we review the quantum computing techniques we will apply.
Considering two different prepotentials, in Section~\ref{sec:results} we present our initial results on dynamical supersymmetry breaking, and also comment on prospects for real-time evolution.

\section{Wess--Zumino Model}\label{sec:wz}
The 1+1-dimensional $\cN = 1$ Wess--Zumino model involves a two-component fermionic field $\psi$ and a bosonic field $\phi$.
It can be considered essentially a supersymmetric extension of $\phi^4$ theory.
Following Refs.~\cite{Beccaria:2001qm, Beccaria:2004pa}, we construct the lattice Hamiltonian $H=Q^2$ on the basis of the discretized supercharge
\begin{equation}
  Q = \frac{1}{\sqrt{a}}\sum_{n=1}^N \left[p_n \psi_{1,n}-\left(\frac{\phi_{n+1}-\phi_{n-1}}{2}+aV(\phi_n)\right)\psi_{2,n}\right],
\end{equation}
for $N$ spatial sites separated by lattice spacing `$a$' (time remains continuous).
Here $V(\phi_n)$ is an arbitrary real `prepotential' that depends on the bosonic field, and $p_n$ is the momentum conjugate to $\phi_n$.
Squaring this supercharge, we find the Hamiltonian
\begin{align}
  H = \sum_n & \left[ \frac{p_n^2}{2a} + \frac{a}{2}\left(\frac{\phi_{n+1}-\phi_{n-1}}{2a}\right)^2+\frac{a}{2}V(\phi_n)^2 + aV(\phi_n)\frac{\phi_{n+1}-\phi_{n-1}}{2a} \right. \nonumber \\
             & \quad \left.+(-1)^nV'(\phi_n)\left(\chi_n^{\dagger}\chi_n-\frac{1}{2}\right) + \frac{1}{2a}\left(\chi_n^{\dagger}\chi_{n+1}+\chi_{n+1}^{\dagger}\chi_n\right) \right], \label{eq:ham}
\end{align}
where we have replaced the two fermion components $\psi_{1,n}$ and $\psi_{2,n}$ with creation and annihilation operators $\chi_n^{\dag}$ and $\chi_n$ defined by
\begin{align}
  \psi_{1,n} & =\frac{1-i(-1)^n}{2i^n}\left(\chi_n^{\dagger}+i\chi_n\right) &
  \psi_{2,n} & =\frac{1+i(-1)^n}{2i^n}\left(\chi_n^{\dagger}-i\chi_n\right).
\end{align}

As mentioned in Section~\ref{sec:intro}, we are interested in dynamical supersymmetry breaking in the 1+1d Wess--Zumino model, which depends on the prepotential $V(\phi_n)$.
Considering polynomial prepotentials of degree $q$, tree-level analyses suggest that supersymmetry should remain preserved when $q$ is odd, but may break spontaneously for even $q$~\cite{Beccaria:2004pa}.
In Sections~\ref{sec:linear} and \ref{sec:quad} we will consider $q = 1$ and $q = 2$, respectively, finding agreement with these expectations.

Of course there have been prior investigations of dynamical supersymmetry breaking both for the 1+1d Wess--Zumino model~\cite{Catterall:2003ae, Wozar:2011gu, Beccaria:2001qm, Beccaria:2003gt, Beccaria:2004pa, Beccaria:2004ds, Steinhauer:2014yaa} as well as for more complicated 1+1d systems including super-Yang--Mills~\cite{Catterall:2017xox}, super-QCD~\cite{Catterall:2015tta} and the supersymmetric Gross--Neveu--Yukawa model~\cite{Fitzpatrick:2019cif}.
A persistent challenge in such studies is the severe sign problem associated with spontaneous supersymmetry breaking~\cite{Bergner:2016sbv, Schaich:2022xgy}, which has motivated the development of fermion loop~\cite{Steinhauer:2014yaa}, tensor network~\cite{Kadoh:2018hqq} and conformal truncation~\cite{Fitzpatrick:2019cif} techniques.

We avoid sign problems by using the VQE to estimate the ground state energy, for now considering sufficiently small systems that the results can be checked through classical diagonalization.
Since $H = Q^2$, the ground state energy $E_0 = \opbraket{\Omega}{H}{\Omega} = |Q\ket{\Omega}\!|^2$ vanishes if and only if the ground state is supersymmetric, $Q\ket{\Omega} = 0$.
Otherwise, if supersymmetry is spontaneously broken, the ground state energy is strictly positive.
While the classical computational costs of diagonalization grow very rapidly as the system size increases, the VQE offers hope of efficient determination of ground state energies using future quantum devices, as we now discuss.

\section{Quantum Computing}\label{sec:qc}
To perform computations on quantum devices, we map the bosonic and fermionic
degrees of freedom to qubit degrees of freedom.  Qubits are physically realized as
two state systems.
This allows for a straightforward mapping for the fermions via the Jordan--Wigner transformation,
\begin{align}
  \chi_n^{\dag} & = \frac{1}{2}\left(X_n-iY_n\right) &
  \chi_n & = \frac{1}{2}\left(X_n+iY_n\right),
\end{align}
where $X_n$ and $Y_n$ represent a Pauli gate acting on the $n$-th qubit.

The bosonic degrees of freedom have an infinite-dimensional Hilbert space at each
lattice site and need to be regulated.
To do this we consider them in the harmonic oscillator basis and impose a hard cutoff on the number \La of allowed modes at each site.
It is worth noting that this explicitly breaks supersymmetry, which will only be exactly restored in the $\Lambda\rightarrow\infty$ limit.
The number of qubits needed to define each $\phi_n$ truncated in this way is $n_q \equiv \lceil\log_2\Lambda\rceil$.
The raising and lowering operators become
\begin{align}
  \ahat_n & = \sum_{l=0}^{\Lambda-2}\sqrt{l+1}\matElem{l}{l+1}, &
  \ahatdag_n & = \sum_{l=0}^{\Lambda-2}\sqrt{l+1}\matElem{l+1}{l}.
\end{align}
The introduction of this cutoff makes the bosonic Hilbert space finite
and we can now perform the mapping to qubits.

We follow the same steps as in Ref.~\cite{Culver:2021rxo}, where more details can be found.
Writing the state $j$ in binary as $j=\sum_{i=0}^{n_q-1}b_i2^i$, we associate each digit with a qubit.  Specific matrix
elements can be converted to their action on qubits using the relations
\begin{align}
  \matElem{0}{1} & =\frac{1}{2}\left(X+iY\right), &
  \matElem{1}{0} & =\frac{1}{2}\left(X-iY\right), \\
  \matElem{0}{0} & =\frac{1}{2}\left(1+Z\right), &
  \matElem{1}{1} & =\frac{1}{2}\left(1-Z\right)
\end{align}
and writing the full matrix element as a tensor product over all of the binary digits:
\begin{equation}
  \matElem{n}{n'} = \otimes_{i=0}^{n_q-1}\matElem{b_i}{b_i'}.
\end{equation}
This completes the mapping of all the degrees of freedom into qubits, which enables the application of quantum algorithms of interest.

An important NISQ-era algorithm for determining whether or not supersymmetry is dynamically broken is the VQE algorithm~\cite{McClean_2016}.
This algorithm outputs an upper bound on the lowest eigenvalue of any matrix
and by using the Hamiltonian as the target matrix we can investigate whether or not the ground state energy is zero.  Specifically we want to test whether or not,
in the $\Lambda\rightarrow\infty$ limit, the VQE estimate of the energy
tends towards zero or to a finite value.  To run the VQE algorithm, we first prepare
some trial wavefunction for the ground state $\psi$ with tunable parameters $\theta_i$.
The energy of this trial state is computed with a quantum circuit and fed into a classical optimization algorithm that adjusts the parameters $\theta_i$ in search of the minimum energy.
This hybrid classical--quantum algorithm will converge to some $E_{\text{var}}$ which is an upper bound for the ground state energy,
\begin{equation}
  E_0\leq E_{\text{var}}=\frac{\opbraket{\psi(\theta_i)}{H}{\psi(\theta_i)}}{\braket{\psi(\theta_i)}{\psi(\theta_i)}}.
\end{equation}

Another important algorithm, more relevant for the long-term prospects of quantum computing,
is the Suzuki--Trotter decomposition of the time-evolution operator, $e^{iHt}$.
This gives us direct access to the real-time evolution of a quantum state acting
under the dynamics of our Hamiltonian.  The continuous time $t$ is broken down into
$N_t$ steps of size $\de = t / N_t$,
\begin{equation}
    e^{-i H t} \ket{\psi} = \left(\exp\left[-i H \de\right]\right)^{N_t} \ket{\psi}.
\end{equation}
Noting that the Hamiltonian will be a sum of $M$ terms, a single Trotter step is
\begin{equation*}
  \ket{\psi(t + \de)} = \exp\left[-i \sum_{j = 1}^M H_j \de\right] \ket{\psi(t)}.
\end{equation*}
For most physical applications $M>1$ and the Baker--Campbell--Hausdorff formula is
used to convert the above line into a product of exponentials of each of the $H_j$.
Unfortunately all of the $H_j$ terms of the Hamiltonian do not commute in general,
thus there will be a tradeoff introduced between the number of gate operations to
perform a single step, and the accuracy to which the final state is obtained.
This error in each step will then be compounded by the time discretization $\de$.
Of course taking a larger number of smaller steps will reduce this error,
but this comes at the cost of more qubit operations which is not ideal in the NISQ era.
We use the default implementation for transpiling the Trotter circuit in Qiskit~\cite{Qiskit}, based on the Hamiltonian in \eq{eq:ham}.
See Ref.~\cite{Ostmeyer:2022lxs} for further discussion of how Trotter decompositions can be optimized for quantum computing.

\section{Results}\label{sec:results}
We now present results for the ground state energy of the Wess--Zumino model and the entangling gate count for a single Trotter step of real-time evolution. 
The ground state energy $E_0$ is important since supersymmetry is preserved if and only if $E_0 = 0$.
We consider a linear prepotential $V(\phi)$ in Section~\ref{sec:linear} and a family of quadratic prepotentials in Section~\ref{sec:quad}, in each case using $N = 2$--$3$ spatial lattice sites and several small values of the bosonic cutoff $\Lambda \leq 16$.
While we can analyze any positive cutoff, in practice \La should be a power of 2 to optimize the correspondence between the bosonic degrees of freedom and the qubit degrees of freedom.
Otherwise our approach would leave part of the quantum computer's Hilbert space unused.
For each calculation we find the ground state energy via classical diagonalization, and see how well this can be reproduced by a modest number of VQE runs (minimum 100).

\subsection{Linear prepotential}\label{sec:linear}
The linear prepotential is just
\begin{equation}
  \label{eq:linear}
  V(\phi_n)=\phi_n,
\end{equation}
and is expected to preserve supersymmetry.
This is the simplest nontrivial $\phi$-dependent prepotential, and keeps the bosonic and fermionic fields from interacting with each other.

\begin{figure}
  \includegraphics[width=0.45\linewidth]{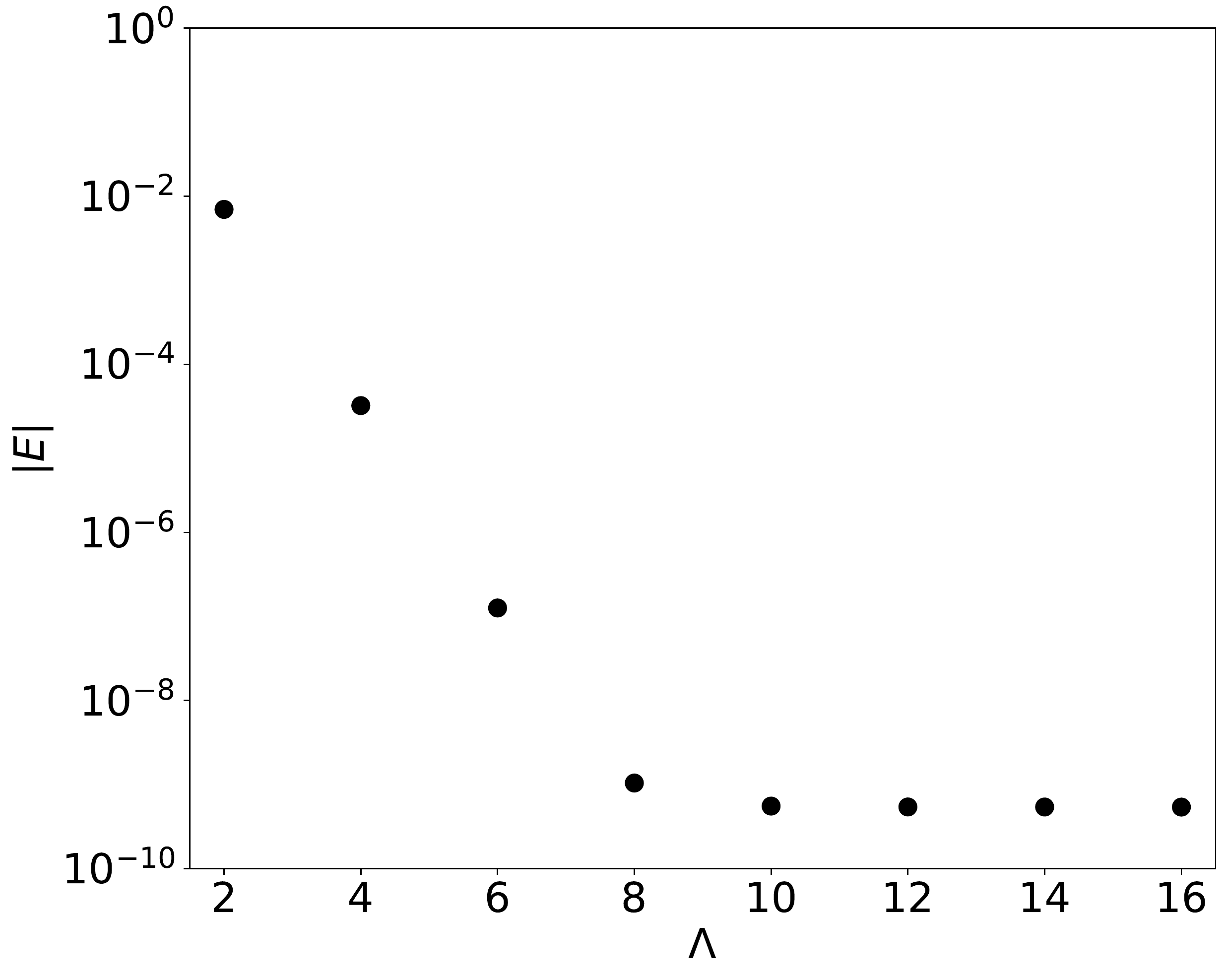}
  \hfill
  \includegraphics[width=0.45\linewidth]{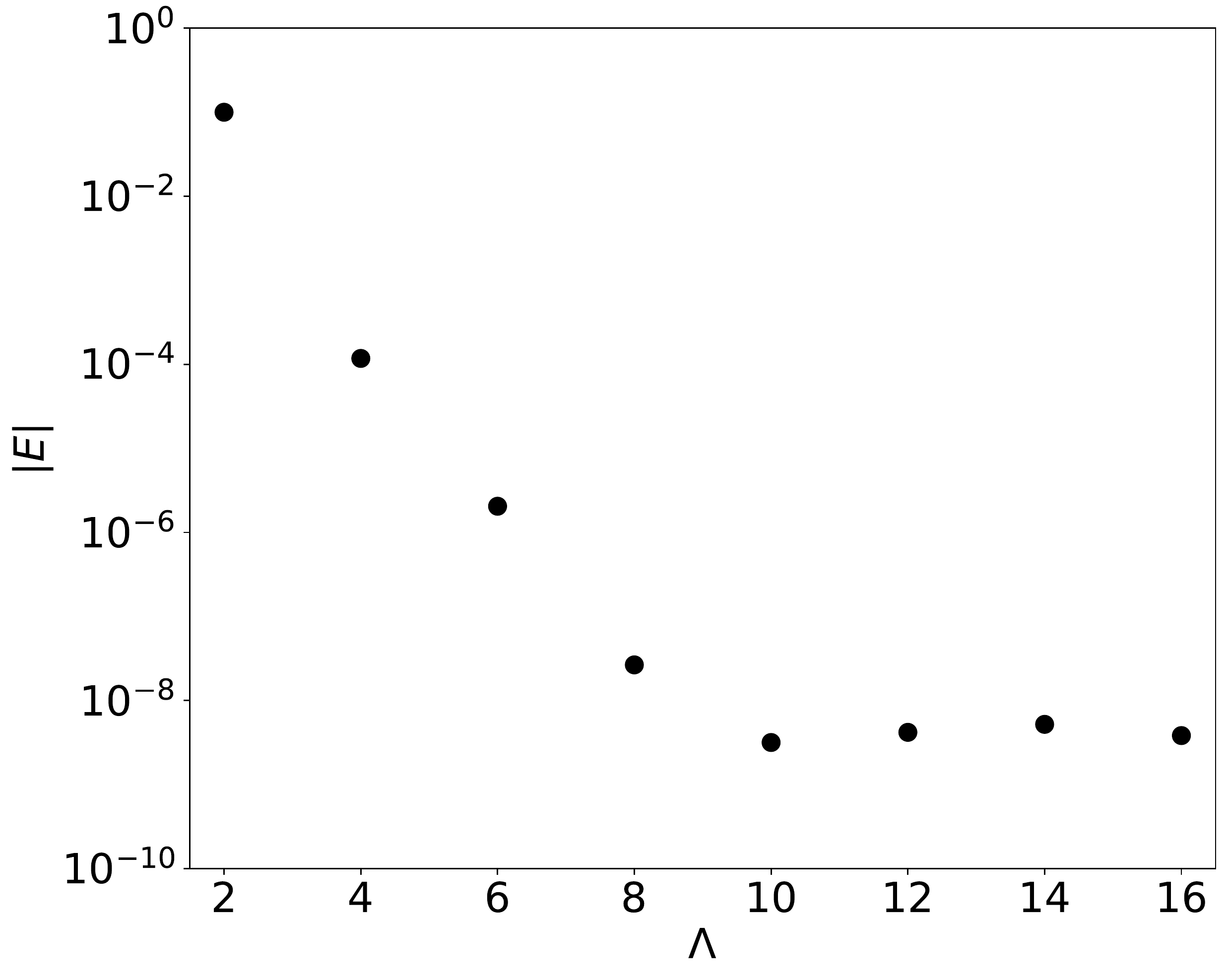}
  \caption{Semi-log plots of the 1+1d Wess--Zumino ground state energy for 2-site~(\textbf{left}) and 3-site~(\textbf{right}) lattices with the linear prepotential \eq{eq:linear}.  As the bosonic cutoff \La goes to infinity the ground state energy goes to zero (up to numerical precision), consistent with the expected preservation of supersymmetry for this prepotential.}
  \label{fig:linear_spectrum}
\end{figure}

\begin{table}
  \caption{\label{table:vqe}Ground state energies from classical diagonalization and the VQE for the linear and quadratic prepotentials discussed in the text.  The linear prepotential \eq{eq:linear} is expected to preserve supersymmetry and have a $\La \to \infty$ ground state energy of zero.  Supersymmetry is expected to break dynamically for the quadratic prepotential \eq{eq:quad}, so long as $c < c_0 \approx -0.5$.}
  \begin{subtable}[t]{0.32\linewidth}
    \centering
    \begin{tabular}{c @{\hspace{1.0\tabcolsep}} c @{\hspace{1.0\tabcolsep}} r @{\hspace{1.0\tabcolsep}} r}
      \hline
      N & $\Lambda$ & Exact & VQE  \\
      \hline\hline
      2 & 2 & 6.97e-03 & 6.97e-03 \\
      - & 4 & 3.22e-05 & 6.61e-05 \\
      - & 8 & 1.04e-09 & 1.08e-01 \\
      \hline
      3 & 2 & -9.97e-02 & -1.28e+00 \\
      - & 4 & 1.17e-04 & 4.99e-01 \\
      \hline
    \end{tabular}
    \caption{\label{table:vqe_a}Linear prepotential}
  \end{subtable}\hfill
  \begin{subtable}[t]{0.32\linewidth}
    \centering
    \begin{tabular}{c @{\hspace{1.0\tabcolsep}} c @{\hspace{1.0\tabcolsep}} r @{\hspace{1.0\tabcolsep}} r}
      \hline
      N & $\Lambda$ & Exact & VQE  \\
      \hline\hline
      2 & 2 & -4.87e-01 & -9.11e-01 \\
      - & 4 & 1.82e-01 & 2.26e-01 \\
      - & 8 & 1.31e-01 & 7.49e-01 \\
      \hline
      3 & 2 & -1.98e-01 & -1.28e+00 \\
      - & 4 & 3.02e-01 & 5.08e-01 \\
      \hline
    \end{tabular}
    \caption{Quadratic prepotential, $c=-0.2$}
  \end{subtable}\hfill
  \begin{subtable}[t]{0.32\linewidth}
    \centering
    \begin{tabular}{c @{\hspace{1.0\tabcolsep}} c @{\hspace{1.0\tabcolsep}} r @{\hspace{1.0\tabcolsep}} r}
      \hline
      N & $\Lambda$ & Exact & VQE  \\
      \hline\hline
      2 & 2 & -4.87e-01 & -9.11e-01 \\
      - & 4 & 1.28e-01 & -1.15e+00 \\
      - & 8 & -1.74e-02 & 6.89e-01 \\
      \hline
      3 & 2 & -1.98e-01 & -1.28e+00 \\
      - & 4 & 2.47e-01 & -1.10e+00 \\
      \hline
    \end{tabular}
    \caption{Quadratic prepotential, $c=-0.8$}
  \end{subtable}
\end{table}

The ground state energy as a function of \La from classical computations of the eigenvalues of $H$ on 2- and 3-site lattices is shown in Fig.~\ref{fig:linear_spectrum}.
In the $\Lambda\rightarrow\infty$ limit the ground state energy goes to zero up to the precision of the solver, which confirms that supersymmetry is preserved as expected.
In Table~\ref{table:vqe_a} we compare the minimum energy obtained from 100 runs of the VQE against these exact results from classical diagonalization.
For the 2-site lattice with $\La \leq 4$ the VQE estimate of the ground state energy behaves appropriately, and exponentially approaches zero.
As an aside, the negative energies shown in Table~\ref{table:vqe} for $\La = 2$ reflect the significant supersymmetry breaking related to such an extreme truncation.
For larger values of the cutoff, and for the larger 3-site lattice, the VQE struggles to converge to the correct order of magnitude.
This provides us with immediate targets we can use in work to improve the performance and reliability of VQE determinations of the ground state energy.

In Table~\ref{table:gate_count_a} we show the entangling CX gate counts for a single Trotter step of real-time evolution for the Wess--Zumino model.
As the bosonic cutoff and the number of sites increase, our time-evolution circuits quickly exceed the capabilities of NISQ hardware.
This motivates ongoing work to optimize time-evolution schemes for the Wess--Zumino model, in order to minimize resource requirements.

\subsection{Quadratic prepotential}\label{sec:quad}
\begin{figure}
  \includegraphics[width=0.45\linewidth]{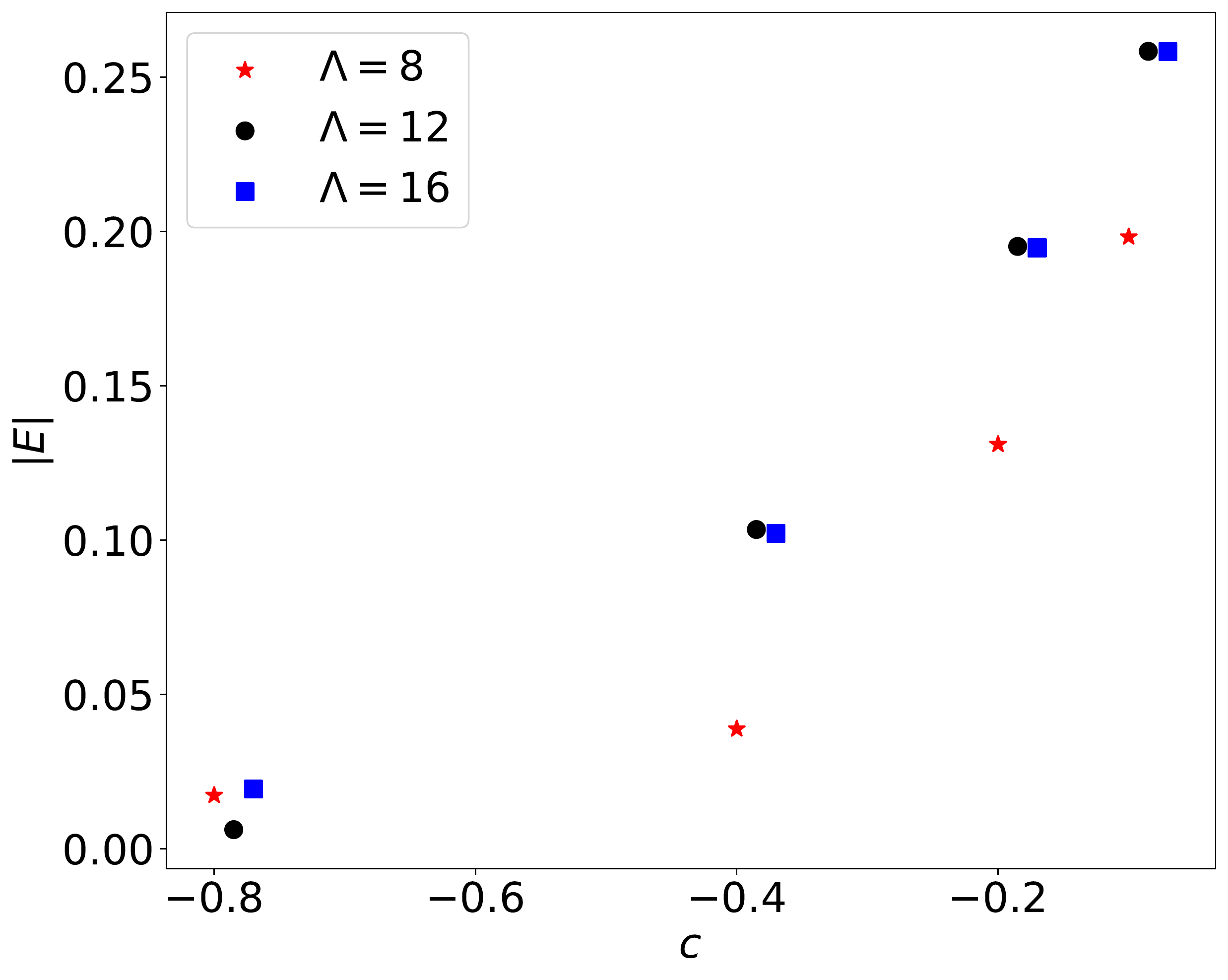}
  \hfill
  \includegraphics[width=0.45\linewidth]{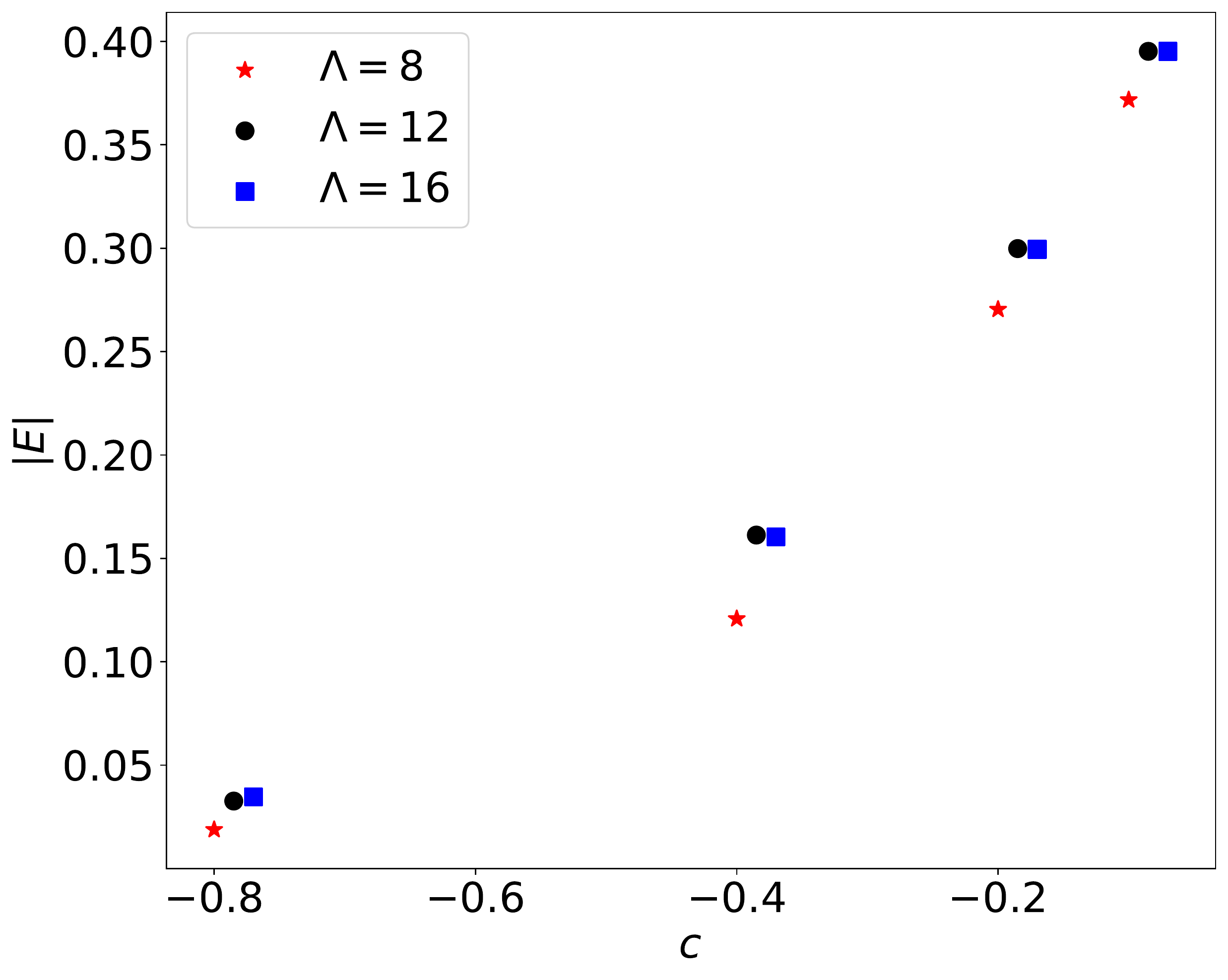}
  \caption{The 1+1d Wess--Zumino ground state energy for 2-site~(\textbf{left}) and 3-site~(\textbf{right}) lattices with various values of the parameter $c$ in the quadratic prepotential of \eq{eq:quad}.  Small horizontal offsets distinguish different values of the cutoff $\La = 8$, $12$ and $16$.  Since non-zero values of $E$ correspond to supersymmetry breaking, the results are consistent with the expected preservation of supersymmetry for sufficiently negative $c < c_0 < 0$ and $\La \to \infty$.}
  \label{fig:quadratic_spectrum}
\end{figure}

To introduce interactions between the bosons and fermions we consider the family of quadratic prepotentials
\begin{equation}
  \label{eq:quad}
  V(\phi_n) = c + \phi_n^2,
\end{equation}
with free parameter $c$, which was studied in Refs.~\cite{Beccaria:2001qm, Beccaria:2004pa, Beccaria:2004ds}.
This prepotential is expected to lead to dynamical supersymmetry breaking for positive $c$, while supersymmetry should be preserved for sufficiently negative $c$ less than a critical value $c_0 < 0$, found to be $c_0 \approx -0.5$ in Refs.~\cite{Beccaria:2004pa, Beccaria:2004ds}.

We can investigate these expectations by computing the ground state energy for a range of $c$.
Results from classical computations of the ground state energy as a function of $c$ are shown in Fig.~\ref{fig:quadratic_spectrum} for several values of \La on 2- and 3-site lattices.
We only show results from $\Lambda\geq 8$ since smaller $\Lambda \leq 4$ produce wildly fluctuating results that are clearly not meaningful.
For both 2- and 3-site lattices the ground state energy converges towards zero as $c\rightarrow -\infty$, confirming that supersymmetry is preserved for sufficiently negative $c$.
For $c\approx 0$ little dependence on the cutoff is visible for $\La \geq 12$.
The results become more sensitive to \La as $c$ becomes more negative, illustrating the challenges that computations in this regime will face on near-term quantum hardware.

This is also reflected in the results for the VQE estimate of the ground state energy presented in Table~\ref{table:vqe} for both $c = -0.2$ and $-0.8$.
In the first case of $c=-0.2$, 100 runs of the VQE suffice to recover the correct order of magnitude of the exact ground-state energy for the 2- and 3-site lattices with $\La \leq 8$ we have considered so far. 
The VQE analysis becomes significantly more challenging for $c=-0.8$.
Since these values of \La are too small to be included in Fig.~\ref{fig:quadratic_spectrum}, for the time being we are focusing our VQE improvement efforts on the linear prepotential discussed in the previous subsection.

\begin{table}
  \caption{\label{table:gate_count}{Entangling CX gate counts for a single Trotter step of the time-evolution operator for the linear and quadratic prepotentials discussed in the text.  For each prepotential, we consider 2- and 3-site lattices, and transpile the circuit for a few values of the bosonic cutoff $\Lambda$.}}
  \begin{subtable}[t]{0.32\linewidth}
    \centering
    \begin{tabular}{c c c}
      \hline
      $N$ & $\Lambda$ & CX Gates\\
      \hline\hline
      2 & 2 & 8 \\
      - & 4 & 252 \\
      - & 8 & 2556 \\\hline
      3 & 2 & 18 \\
      - & 4 & 5728 \\
      \hline
    \end{tabular}
    \caption{\label{table:gate_count_a}Linear prepotential}
  \end{subtable}\hfill
  \begin{subtable}[t]{0.32\linewidth}
    \centering
    \begin{tabular}{c c c}
      \hline
      $N$ & $\Lambda$ & CX Gates\\
      \hline\hline
      2 & 2 & 14 \\
      - & 4 & 754 \\
      - & 8 & 7822 \\ \hline
      3 & 2 & 30 \\
      - & 4 & 2788 \\
      \hline
    \end{tabular}
    \caption{Quadratic prepotential, $c=-0.2$}
  \end{subtable}\hfill
  \begin{subtable}[t]{0.32\linewidth}
    \centering
    \begin{tabular}{c c c}
      \hline
      $N$ & $\Lambda$ & CX Gates\\
      \hline\hline
      2 & 2 & 14 \\
      - & 4 & 718 \\
      - & 8 & 7858 \\ \hline
      3 & 2 & 30 \\
      - & 4 & 2730 \\
      \hline
    \end{tabular}
    \caption{Quadratic prepotential, $c=-0.8$}
  \end{subtable}
\end{table}

Finally, in Table~\ref{table:gate_count} we again provide entangling gate counts for a single Trotter step.
The results are qualitatively similar to those for the linear prepotential.
For $\La > 2$, the interactions between the bosons and fermions increase CX gate requirements by roughly a factor of 3. 
The mild dependence on the value of $c$ is related to the realization of $c$-dependent terms in the Hamiltonian in terms of the default basis gate set.

\section{Conclusion}
We have presented initial results from our ongoing work using quantum computing to study the Wess--Zumino model in 1+1 dimensions.
We use a Hamiltonian lattice regularization of the theory to analyze spontaneous supersymmetry breaking for prepotentials with linear or quadratic dependence on the bosonic field.
The preservation or breaking of supersymmetry is determined by the ground state energy, which we analyze with the VQE algorithm for systems with 2 or 3 spatial sites and small bosonic cutoffs $\Lambda \leq 16$ --- small enough to check our results through classical diagonalization.

Despite the small size of these systems, our results are consistent with the expected preservation of supersymmetry for the linear prepotential, and show clear spontaneous supersymmetry breaking for a quadratic prepotential with $c = -0.2$.
However, our VQE algorithm struggles to converge for large values of the bosonic cutoff, motivating ongoing work to improve the performance and reliability of this approach to exploring spontaneous supersymmetry breaking.

We also consider prospects for studying the real-time dynamics of the Wess--Zumino model, based on a straightforward transpilation of the time-evolution operator corresponding to the Hamiltonian \eq{eq:ham}.
Even for 2-site lattices, the number of gate operations in the resulting time-evolution circuits exceed the capabilities of NISQ hardware for any reasonable cutoff $\La > 2$.
Here as well we hope to improve this situation by more carefully considering time-evolution schemes in search of optimal use of quantum resources. 

\vspace{12 pt}
\noindent \textsc{Acknowledgments:} We thank Johann Ostmeyer for discussions of Trotter schemes.
This work was supported by UK Research and Innovation Future Leader Fellowship {MR/S015418/1} and STFC grant {ST/T000988/1}.

\bibliographystyle{JHEP}
\bibliography{lattice22}

\end{document}